# Mixed anion control of enhanced negative thermal expansion in the oxysulfide of PbTiO$_3$

Zhao Pan,[1,2,*] Zhengli Liang,[3] Xiao Wang,[1] Yue-Wen Fang,[4,5,*] Xubin Ye,[1] Zhehong Liu,[1] Takumi Nishikubo,[6,2] Yuki Sakai,[6,2] Xi Shen,[1] Qiumin Liu,[2] Shogo Kawaguchi,[7] Fei Zhan,[8] Longlong Fan,[8] Yong-Yang Wang,[8] Chen-Yan Ma,[8] Xingxing Jiang,[3] Zheshuai Lin,[3] Richeng Yu,[1] Xianran Xing,[9,*] Masaki Azuma,[2,6] and Youwen Long[1,10,*]

[1]*Beijing National Laboratory for Condensed Matter Physics, Institute of Physics, Chinese Academy of Sciences, Beijing 100190, China*

[2]*Laboratory for Materials and Structures, Tokyo Institute of Technology, 4259 Nagatsuta, Midori, Yokohama, 226-8503, Japan*

[3]*Center for Crystal R&D, Key Laboratory of Functional Crystals and Laser Technology, Technical Institute of Physics and Chemistry, Chinese Academy of Sciences, Beijing 100190, China*

[4]*Fisika Aplikatua Saila, Gipuzkoako Ingeniaritza Eskola, University of the Basque Country (UPV/EHU), Europa Plaza 1, 20018 Donostia/San Sebastián, Spain*

[5]*Centro de Física de Materiales (CSIC-UPV/EHU), Manuel de Lardizabal Pasealekua 5, 20018 Donostia/San Sebastián, Spain*

[6]*Kanagawa Institute of Industrial Science and Technology (KISTEC), 705-1 Shimoimaizumi, Ebina, Kanagawa 243-0435, Japan*

[7]*Research and Utilization Division, Japan Synchrotron Radiation Research Institute (JASRI), SPring-8, 1-1-1 Kouto, Sayo-cho, Sayo-gun, Hyōgo 679-5198, Japan*

[8]*Synchrotron Radiation Laboratory, Institute of High Energy Physics, Chinese Academy of Sciences, Beijing 100049, China*

[9]*Beijing Advanced Innovation Center for Materials Genome Engineering and Department of Physical Chemistry, University of Science and Technology Beijing, Beijing 100083, China*

[10]*Songshan Lake Materials Laboratory, Dongguan, Guangdong 523808, China*

[†]*Electronic supplementary information (ESI) available.*

Corresponding author: zhaopan@iphy.ac.cn; yuewen.fang@ehu.eus; xing@ustb.edu.cn; ywlong@iphy.ac.cn

**New concepts**

PbTiO$_3$ (PT), a typical perovskite-type ($AB$O$_3$) ferroelectric, which also exhibits unique negative thermal expansion (NTE) property. Materials with strong NTE are very important, since they can be used to tailor the overall coefficient of thermal expansion (CTE) of materials, or even realize the favorable zero thermal expansion. While much attention has been paid to modify the NTE property of PT by replacing A-site Pb and/or B-site Ti cations, the effect of anion-site substitution on the NTE of PT remains unknown. Therefore, a feasible mixed anion control of thermal expansion of PT was proposed. Unusual enhanced NTE over a wide temperature was achieved in the oxysulfide of PbTiO$_3$, which was evidenced by a comprehensive experimental and theoretical studies. The anion-mediated enhanced NTE demonstrates a new technique for realizing large NTE in PT-based perovskites, providing a new way for the design of high-performance NTE functional materials.

**Abstract**

The rare physical property of negative thermal expansion (NTE) is intriguing because materials with large NTE over a wide temperature range can serve as high-performance thermal expansion compensators. However, applications of NTE are hindered by the fact that most of the available NTE materials show small magnitudes of NTE, and/or NTE occurs only in a narrow temperature range. Herein, for the first time, we investigated the effect of anion substitution instead of general Pb/Ti-site substitutions on the thermal expansion properties of a typical ferroelectric NTE material, PbTiO$_3$. Intriguingly, the substitution of S for O in PbTiO$_3$ further increases the tetragonality of PbTiO$_3$. Consequently, an unusually enhanced NTE with an average volumetric coefficient of thermal expansion $\bar{\alpha}_V$ = -2.50 × 10$^{-5}$/K was achieved over a wide temperature range (300 – 790 K), which is contrasted to that of pristine PbTiO$_3$ ($\bar{\alpha}_V$ = -1.99 × 10$^{-5}$/K, RT – 763 K). The intensified NTE is attributed to the enhanced hybridization between Pb/Ti and O/S atoms by the substitution of S, as evidenced by our theoretical investigations. We therefore demonstrate a new technique for introducing mixed anions to achieve large NTE over a wide temperature range in PbTiO$_3$-based ferroelectrics.

## 1. Introduction

Controlling thermal expansion is an important topic in fundamental research and a critical issue in practical applications. Particularly, in high-precision applications subject to large temperature fluctuations, such as electronic devices and optical instruments.[1-5] The discovery of negative thermal expansion (NTE) materials, in which the volume shrinks instead of expanding upon heating in a certain temperature range, offers a promising opportunity to tailor the coefficient of thermal expansion (CTE) of the materials. Such materials can be combined with normal positive thermal expansion (PTE) materials to form composites with controllable positive, negative, or even zero thermal expansion (ZTE). In the past decades, a wide variety of materials have been reported to exhibit NTE, including oxides,[6-14] alloys,[15-20] nitrides,[21-23] fluorides,[24-28] cyanides,[29-31] and reduced ruthenates.[32-34] Although the control of thermal expansion is important, such behavior can be difficult to achieve. Indeed, most of the present NTE materials show small magnitudes of NTE, and NTE usually occurs in a narrow temperature range, which limits the applications of such NTE materials. Therefore, achieving a large NTE over a wide temperature range remains a great challenge.

Among the currently available NTE materials, PbTiO$_3$ (PT)-based ferroelectric NTE materials are particularly interesting. PT is a typical perovskite-type ($AB$O$_3$) ferroelectric with a large tetragonality ($c/a$) of 1.064.[35] In addition to ferroelectricity, PT exhibits a unique NTE in the perovskite family.[7] The unit cell volume of PT contracts over a wide temperature range from room temperature to its Curie temperature ($T_C$ = 763 K) in the tetragonal ferroelectric phase, with an average intrinsic bulk CTE of $\bar{\alpha}_V$ = -1.99 × 10$^{-5}$ K$^{-1}$.[36] Owing to the advantage of a flexible perovskite structure, the NTE property of PT can be efficiently controlled by varying the substitutions at the A-site Pb and B-site Ti atoms.[3] Because the volume shrinkage of PT only occurs along the polar $c$ axis of the tetragonal phase, it is proposed that the increase in $c/a$ could induce a larger NTE. Indeed, enhanced NTEs were obtained in PT-based ferroelectrics with improved tetragonality.[37-40] Nevertheless, most substitutions in PT result in reduced tetragonality and weakened NTE (see panel I in Fig. 1), and only a few PT-based ferroelectrics exhibit unusually improved tetragonality and enhanced NTE compared with pristine PT (see panel II in Fig. 1).

To date, all modifications of the NTE property of PT have focused on the $A$- and/or $B$-site cation substitutions, while no studies have reported on the effect of substitutions at the anion O site on the NTE of PT. One possible reason for this could be the difficulty in preparing PT-based mixed-anion compounds. However, the differences in the valences, electronegativities, and ionic radii of heteroanions are expected to affect the crystal and electronic structure of PT, and therefore, modify the NTE behavior.[41] In the present study, we successfully synthesized the oxysulfide of $PbTiO_3$ using a unique high-pressure and high-temperature (HPHT) method. Intriguingly, improved tetragonality accompanied by an unusually enhanced NTE was obtained, which indicates that the introduction of anions could be a new and effective way to achieve a large NTE in PT-based ferroelectric NTE materials.

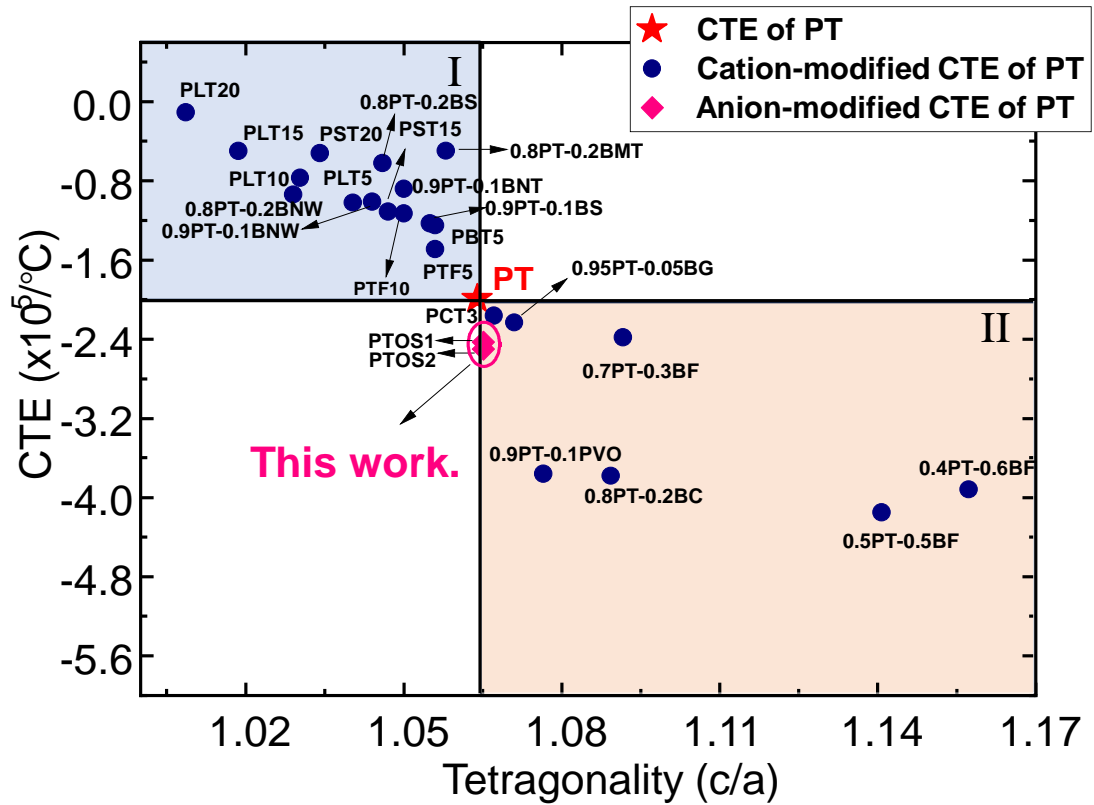

**Fig. 1** Relationship between tetragonality ($c/a$) and coefficient of thermal expansion (CTE) in PT-based ferroelectrics. Herein, PLT100$x$ is the abbreviation of $Pb_{1-x}La_xTiO_3$, PST100$x$ indicates $Pb_{1-x}Sr_xTiO_3$, PTF100$x$ stands for $Pb(Ti_{1-x}Fe_x)O_3$, PBT100$x$ represents $(Pb_{1-x}Bi_x)TiO_3$, (1-$x$)PT-$x$BNT indicates (1-$x$)$PbTiO_3$-$x$Bi(Ni$_{1/2}$Ti$_{1/2}$)$O_3$, PCT100$x$ stands for $Pb_{1-x}Cd_xTiO_3$, (1-$x$)PT-$x$PV represents (1-$x$)$PbTiO_3$-$x$PbVO$_3$, (1-$x$)PT-$x$BF indicates (1-$x$)$PbTiO_3$-$x$BiFeO$_3$, and (1-$x$)PT-$x$BC indicates (1-$x$)$PbTiO_3$-$x$BiCoO$_3$, respectively.[3]

## 2. Results and discussion

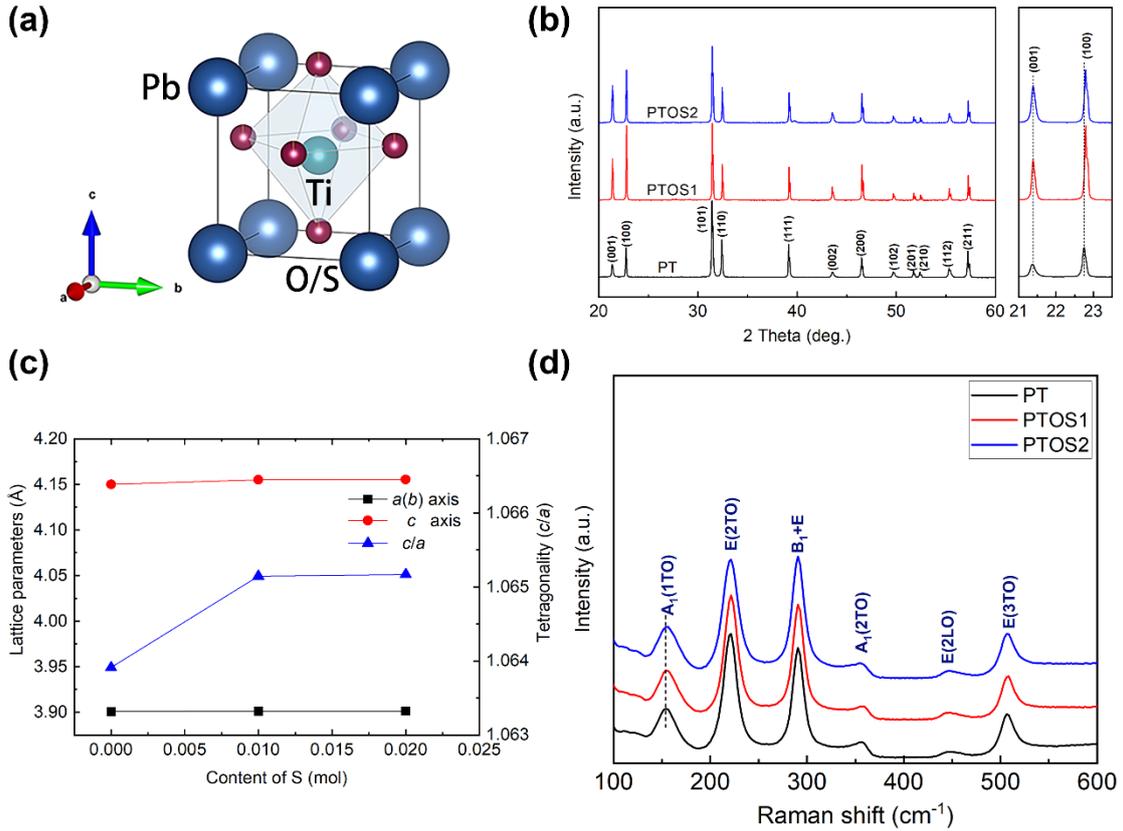

**Figure 2**. Crystal structure and lattice parameters of $PbTiO_{3-x}S_x$. a) Schematic of the perovskite structure with a *P4mm* symmetry. b) Evolution of XRD patterns, c) tetragonality (*c/a*) (The error bars are much smaller than the symbols), and d) Raman spectra of the $PbTiO_{3-x}S_x$ ($x$ = 0, 0.01, and 0.02) compounds as a function of sulfide concentration at room temperature.

According to the X-ray diffraction (XRD) patterns of the $PbTiO_{3-x}S_x$ compounds (Fig. 2b), all the samples are of high quality with a perovskite structure (Fig. 2a). Note that impurities appeared for *x* values higher than 0.02, indicating the solid solubility limit was reached under those synthesis conditions. Although the substitution content of sulfur was low, the S distribution was very uniform in the lattice according to the elemental distribution analysis (Supplemental Fig. 1). Additionally, according to the XANES spectra, the main absorption at the S-*K* edge of PTOS2 is located at 2472.8 eV (Supplemental Fig. 2), which is identical to results reported for $S^{2-}$,[42] thus suggesting the existing of $S^{2-}$ in the PTOS compounds. Note that all the investigated samples can be indexed to the tetragonal symmetry with the space group of *P4mm*. The detailed structural parameters for the $PbTiO_{3-x}S_x$ compounds were refined using Synchrotron X-ray diffraction (SXRD) data and are listed in Supplemental Figs. (3, 4) and Supplementary Table 1. With the substitution of S for O, there is almost no shift for the (001) peak, indicating the *c* axis is stable. However, the (100) peak exhibits an apparent shift toward the higher angle region as a function of S (see the amplified part around 2*θ*

~ 21-23° of Fig. 2b.), which suggests the contraction of the $a$ axis. As a result, the $c/a$ ratio slightly increased from 1.064 for pristine PT to ~1.065 for PTOS1 and PTOS2 (Fig. 2c), which implies the enhanced tetragonality by substitution of S. The enhanced tetragonality can be attributed to the strong Pb-O/S and Ti-O/S hybridizations, which will be discussed in detail later.

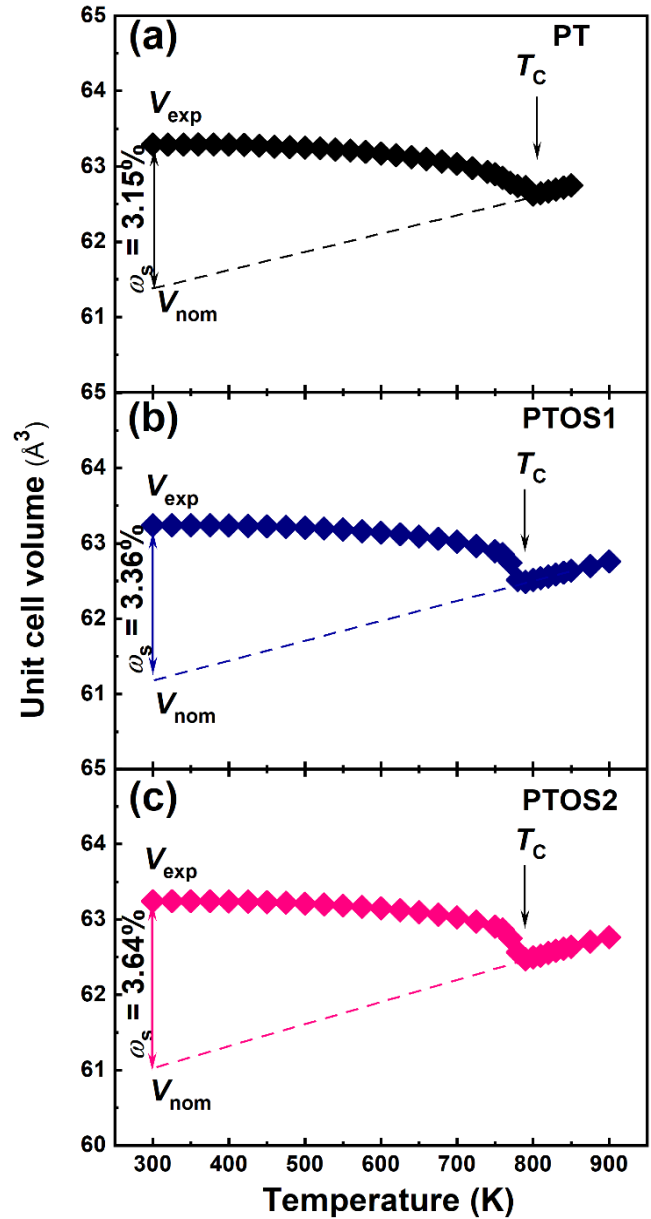

**Figure 3**. Thermal expansion property. Temperature dependence of the unit cell volume of the PbTiO$_{3-x}$S$_x$ compounds. a) $x = 0$, b) $x = 0.01$, and c) $x = 0.02$. The schematic diagram of spontaneous volume ferroelectrostriction ($\omega_S$) and $T_C$ are also presented. Note that the error bars are much smaller than the symbols.

Based on the lattice dynamic theory, the vibrational soft modes are proposed to be closely related to the ferroelectric phase transition. It is suggested that the frequency of the A$_1$(1TO) soft mode is proportional to the order parameter of spontaneous polarization ($P_S$) as it represents the displacement of the $BO_6$ octahedron relative to $A$-site atoms in perovskite-type ferroelectric materials. Therefore, the $A_1$(1TO) soft mode can indicate the variation of $P_S$ in PT-based perovskites both with normally reduced $c/a$ and abnormally enhanced $c/a$. For example, an enhanced $c/a$ accompanied with hardened $A_1$(1TO) was observed in $Pb_{1-x}Cd_xTiO_3$,[37] which an opposite tendency of reduced $c/a$ and softened $A_1$(1TO) was observed in $Pb_{1-x}Sr_xTiO_3$.[43] The Raman spectra of the present PTOS compounds are shown in Fig. 2d. Intriguingly, the Raman active modes of $A_1$(1TO) abnormally show slightly shift to higher frequency with increasing S content, which is consistent with the enhanced $c/a$. Note that this is the first observation of mixed anion control of enhanced tetragonality in $PbTiO_3$-based perovskites.

As mentioned before, in PT-based ferroelectric NTE materials, the enhanced $c/a$ could be related to a large ferroelectric volume effect, that is, a large NTE.[40,44] Generally, a large $c/a$ of PT-based ferroelectrics indicates a large lattice distortion and can result in a large volume shrinkage. Consequently, a large NTE can usually be obtained in PT-based systems with an enhanced $c/a$ compared to that of pristine PT.[3] To study the thermal expansion property of the present PTOS100$x$ compounds with increased tetragonality, temperature dependence of the SXRD experiment was performed for the PTOS1 and PTOS2 samples. The unit cell volume as a function of temperature was extracted by structural refinement based on SXRD data. Pristine PT exhibits a nonlinear and strong NTE from RT to 800 K with an average volumetric CTE of $\bar{\alpha}_V$ = -2.11 × 10$^{-5}$/K (Fig. 3a). Notably, the difference in CTE of PT between the present study and the previous report could be attributed to the different accuracy of sample temperatures during the high-temperature XRD measurements. With the substitution of S for O, the PTOS1 compound exhibits a nonlinear but stronger NTE from room temperature (RT) to $T_C$ ~ 790 K with an average volumetric CTE of $\bar{\alpha}_V$ = -2.43 × 10$^{-5}$/K (Fig. 3b). With a further increase in the content of S for the PTOS2 compound relative to PTOS1, the magnitude of NTE continued to increase, and a further enhanced NTE with an average volumetric CTE of $\bar{\alpha}_V$ = -2.50 × 10$^{-5}$/K was achieved over a wide temperature range from RT to 790 K (Fig. 3c) for PTOS2. Herein, we successfully achieved enhanced NTE of PT by anion substitution for the first time, which provides a new strategy for obtaining large NTE in PT-based ferroelectric NTE materials.

In comparison, the NTE in the present PTOS is much stronger than many PT-based ferroelectrics, with enhanced NTE compared to the pristine PT, such as $Pb_{0.97}Cd_{0.03}TiO_3$ ($c/a$ = 1.066, $\bar{\alpha}_V$ = -2.16 × 10$^{-5}$/K, 300 ~ 743 K)[37] and 0.95PT-0.05BiGaO$_3$ ($c/a$ = 1.071, $\bar{\alpha}_V$ = -2.23 × 10$^{-5}$/K, 300 ~ 773 K).[38] Additionally, the present NTE is even comparable to typical framework NTE materials, such as $ZrW_2O_8$ ($\bar{\alpha}_V$ = -2.73 × 10$^{-5}$/K, 0.3 – 1050 K) and $Fe[Co(CN)_6]$ ($\bar{\alpha}_V$ = -4.41 × 10$^{-6}$/K, 4.2 – 300K),[6,29] which were usually reported to show strong NTE.

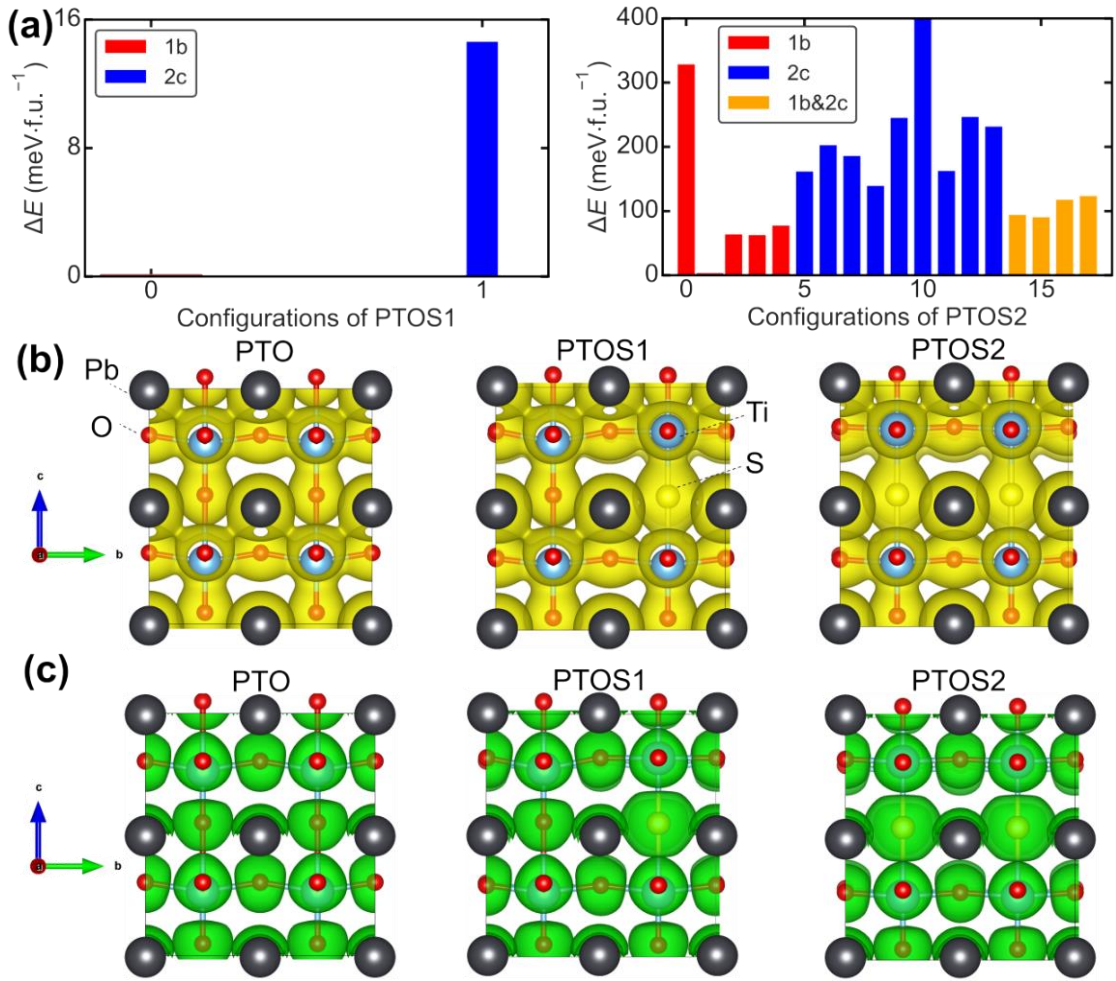

**Figure 4.** Electron density. a) Energetic profile for PTOS1 and PTOS2. The horizontal index indicates the 'ID' of the doped configurations in our calculations. The lowest energy configuration is chosen as the reference and is set to zero. b) Charge density maps of PTO, PTOS1 and PTOS2. c) Electron localization functions of PTO, PTOS1 and PTOS2, in which the isosurface is set to 0.5.

According to vast experimental and theoretical studies of PT-based ferroelectric NTE materials, ferroelectric behavior plays an important role in the NTE of PT-based ferroelectrics.[3,45] A new physical concept of spontaneous volume ferroelectrostriction (SVFS, $\omega_S$) was recently proposed to quantitatively elucidate the effect of ferroelectricity on the anomalous volume change in the tetragonal ferroelectric phase of PT-based ferroelectrics, where the baseline of the description of SVFS can be estimated for the contribution purely by the thermal expansion of phonon vibration.[46] The SVFS effect can be defined by the following equation,

$$\omega_S = \frac{V_{exp} - V_{nm}}{V_{nm}} \times 100\%$$

where $V_{exp}$ and $V_{nm}$ indicate the experimental and nominal unit cell volumes, respectively. Here, $V_{nm}$ can be estimated by extrapolation from the paraelectric phase to the ferroelectric phase. Note that a large value of $\omega_S$ indicates a strong ferroelectrovolume effect and an enhanced NTE, whereas a small value indicates a weak NTE. The $\omega_S$ values were 3.36% and 3.64% for the PTOS1 and PTOS2 compounds (Fig. 3), respectively, which are much higher than that of PT (3.1%).[3] The above results are consistent with the enhanced NTE observed in PTOS1 and PTOS2.

According to the previous studies,[47,48] the hybridization between the cations and anions is crucial to the NET in PT-based ferroelectrics. To understand the hybridization, we have used first-principles calculations to understand the effect of sulfur doping on lead titanate. Because the concentration in the actual PbTiO$_{3-x}$S$_x$ samples is relatively small ($x$ = 0.01 and 0.02), it is not practical in standard density functional theory (DFT) calculations using the supercell method. To implement the modeling, we substitute one (two) sulfur atom(s) for one (two) oxygen atom(s) to approximate PTOS1 and PTOS2 in a 2×2×2 40-atom supercell. In pristine PbTiO$_3$, there are two different Wyckoff sties 1b and 2c. To find the lowest-energy doped structure, we performed total energy calculations for all the possible structure configurations of PTOS1 and PTOS2. The energetic profile is shown in Fig. 4a. We find that the both PTOS1 and PTOS2 prefer the S doping at the 1b site energetically. To simplify the analysis, we just studied the electronic properties of the two lowest energy state of PTOS1 and PTOS2. Fig. 4b shows the comparison of PT, PTOS1 and PTOS2. Upon doping, the hybridization between in-plane Ti and O atoms is slightly affected; on the contrary, the overlap of the large volume of the isosurface of S and its neighboring out-of-plan Ti indicate the hybridization between S and Ti is strong along [001] axis. By replacing O with S, we find the hybridization of the 1b anion site and the neighboring Pb is also enhanced. The above phenomena can be attributed to that sulfur is more delocalized than oxygen. This delocalization can be understood because oxygen is more electronegative than sulfur and has a stronger tendency to attract electrons towards their nuclei. In Fig. 4c, we also show the electron localization functions at the isosurface of 0.5. This also implies that the sulfur doping introduced more delocalized electronic states, that enhances the hybridization Ti and S, as well as the hybridization between the dopants and the neighboring Pb. The evolution of the hybridization observed in the DFT calculations may be responsible to the enhanced NTE in the sulfur doped $PbTiO_3$.

The issue of achieving large NTE is vital; however, it remains challenging. The discovery of NTE materials provides a promising possibility because NTE materials can be used to compensate for general PTE by forming composites with NTE materials. Therefore, materials with controllable CTE can be developed, which is very important for practical applications. As the two key factors for the application of NTE materials, the magnitude of the NTE and the NTE operation window are critical. However, most currently available NTE materials show small magnitudes of NTE, and NTE occurs only in a narrow temperature range. The present work provides a new and effective route for exploring an enhanced NTE over a wide temperature range in PT-based ferroelectrics by introducing anion substitution at the O site. Under these guidelines, we also prepared other new mixed-anion PT-based compounds. A novel thermal expansion property is also expected. Thus, the introduction of anions is likely to be an effective method for controlling thermal expansion in currently known NTE materials with flexible structures.

## 3. Conclusion

In summary, we successfully prepared the oxysulfide of $PbTiO_3$ using a high-pressure and high-temperature method, to study the effect of anion substitution on the thermal expansion property. We discovered that the tetragonality of pristine PT could be increased even with a small amount of S substitution for O. As a result, an enhanced NTE with an average volumetric CTE of $\bar{\alpha}_V = -2.50 \times 10^{-5}/K$ was achieved over a wide temperature range from RT up to 790 K, which is much stronger than that of pristine PT. Furthermore, first-principles calculations indicate that the enhanced tetragonality in the present PTOS system is attributed to the enhanced hybridization between Pb and O/S by the substitution of S. This work demonstrates a new technique for controlling the CTE of PT-based ferroelectric NTE materials, which can also be applied to other NTE materials with flexible structures.

## 4. Experimental Section

**Sample Preparation**: A series of the $PbTiO_{3-x}S_x$ ($x = 0 \sim 0.02$, hereafter abbreviated as

PTOS100$x$) compounds were prepared by the high-pressure and high-temperature (HPHT) method. The high-purity raw materials of PbO, PbS, and TiO$_2$ were thoroughly mixed according to the stoichiometric ration. The mixtures were then sealed in a platinum capsule with a diameter of 3.6 mm and a height of 5.0 mm, followed by the HPHT treatment at 8 GPa and 1200 °C for 60 min using a cubic anvil-type apparatus.

**Crystal Structure Analysis**: The crystal structures for all the samples have been characterized by means of the laboratory X-ray powder diffraction (XRD) performed on a Brucker D8 ADVANCE diffractometer. Temperature dependence of the synchrotron X-ray diffraction (SXRD) data were collected at the BL02B2 beamline of SPring-8 with the light wavelength of 0.42 Å.[49] The detailed structure analysis was performed with the same initial model as that of PbTiO$_3$ (space group *P4mm*, No. 99) by the Rietveld method using the *FullProf* software.[50] The microstructural and elemental results were acquired on transmission electron microscopes (TEMs) (ARM200F with a field-emission gun (FEG), and JEM-2100 Plus) operating at 200 kV. Raman scattering spectra were collected on a MonoVista CRS+ 500 spectrometer (Spectroscopy & Imaging, Germany). The X-ray absorption near edge spectroscopy (XANES) at the S-$K$ edge were performed at beamline 4B7A of Beijing Synchrotron Radiation Facility (BSRF) via fluorescence (FY) mode. The S and PbS$^{2-}$ were also measured as S$^0$ and S$^{2-}$ references, respectively.

**Electronic Structure Calculation**: The structural and electronic properties of lead titanate and the sulfur doped lead titanate are studied by density functional theory (DFT) methods implemented in the Vienna ab initio simulation package (VASP).[51,52] The GGA-type PBE exchange functional[53] was used with the cutoff energy of 500 eV. All the doped and pristine lead titanate models were simulated using the 2×2×2 40-atom supercell. PTOS1 and PTOS2 were approximately simulated by substituting one sulfur and two sulfur atoms in the supercells, respectively. The cell parameters of PT, PTOS1, and PTOS2 from the experiment were employed in the DFT calculations and were fixed throughout the calculations accordingly. The coordinates, starting from the experimental data, were fully relaxed in calculations to ensure that the energy criterion of 10$^{-6}$ eV and the force criterion of 0.005 eV per angstrom were satisfied.

**Data and materials availability**

All data needed to evaluate the conclusions in the paper are present in the paper and/or the ESI.†

**Author Contributions**

Z.P. conceived the project. Z.P. synthesized the samples. Z.P., X.W., X. B. Y., L.F., Y.Y.W., and C.Y.M. performed the XANES measurements. Y. -W. F., Z.L., X.J., and Z.L. performed the theoretical calculations. Z.P., T.N., Y. S., Q.L., and S.K. performed the synchrotron X-ray diffraction experiments. X.S. and R.Y. measured and analyzed the TEM data. Z.P. wrote the manuscript. All authors discussed the results and commented on the manuscript. Z.P., M.A., X. R. X., and Y.L. guided the projects.

**Conflict of Interest**

The authors declare that they have no competing interests.

**Acknowledgements**

This work was supported by the National Key R&D Program of China (Grant No. 2021YFA1400300), the National Natural Science Foundation of China (22271309, 12304268, 11934017, 11921004, and 12261131499), the Grants-in-Aid for Scientific Research (JP18H05208 and JP19H05625) from the Japan Society for the Promotion of Science (JSPS), the Beijing Natural Science Foundation (Grant No. Z200007), the Chinese Academy of Sciences (Grant No. XDB33000000), and the Kanagawa Institute of Industrial Science and Technology. The synchrotron X-ray powder diffraction experiments were performed at SPring-8 with the approval of the Japan Synchrotron Radiation Research Institute (2023B1575 and 2024A1506). XANES at the S-*K* edge were performed at beamline 4B7A of Beijing Synchrotron Radiation Facility (BSRF).